\documentclass[onecolumn,showpacs,floatfix,preprintnumbers,amsmath,amssymb]{revtex4}

\usepackage{graphicx}
\usepackage{psfig}
\usepackage{dcolumn}
\usepackage{bm}

\begin{document}

\title{
Reorientation in Antiferromagnetic Multilayers:\\ 
Spin-Flop Transition and Surface Effects}
\author{U.K. R\"o\ss ler$^{\mathrm{(a)}}$}
\thanks{Corresponding author
IFW Dresden,
Postfach 270116, D--01171 Dresden, Germany. 
Tel.: +49-351-4659-542; Fax: +49-351-4659-537
}
\email{u.roessler@ifw-dresden.de}

\author{A.N.\ Bogdanov$^{\mathrm{(a,b)}}$}
\email{bogdanov@kinetic.ac.donetsk.ua}

\affiliation{
(a) Leibniz-Institut f{\"u}r Festk{\"o}rper- 
und Werkstoffforschung Dresden\\
Postfach 270116,
D--01171 Dresden, Germany
}%

\affiliation{
(b) Donetsk Institute for Physics and Technology \\
83114 Donetsk, Ukraine
}%

\date{\today}

\begin{abstract}
{
Nanoscale superlattices with uniaxial
ferromagnetic layers antiferromagnetically coupled
through non-magnetic spacers are recently used
as components of magnetoresistive and recording devices.
In the last years intensive experimental 
investigations of these artificial antiferromagnets 
have revealed  a  large variety of surface 
induced reorientational effects 
and other remarkable phenomena  
unknown in other magnetic materials.
In this paper we review and generalize 
theoretical results,
which enable a consistent description 
of the complex magnetization processes
in antiferromagnetic multilayers,
and we explain the responsible physical mechanism.
The general structure of phase diagrams 
for magnetic states in these systems is discussed.
In particular, our results resolve 
the long standing problem of 
a ``surface spin-flop'' in antiferromagnetic layers.
This explains the different 
appearance of field-driven reorientation
transitions in systems like Fe/Cr (001) and (211) superlattices, 
and in [CoPt]/Ru multilayers 
with strong perpendicular anisotropy.
}

\end{abstract}


         
\maketitle

%
%
\textbf{1. Introduction. }

Since the discovery of antiferromagnetic
interlayer exchange \cite{Gruenberg87},
a large variety of magnetic nanostructures 
consisting of stacks of ferromagnetic layers
with antiferromagnetic coupling via 
spacers has been synthesized.
For recent investigations on such superlattices,
see \cite{Wang94,AFM001, Felcher02, Nagy02, Hellwig03} 
and references in \cite{PRB04,PRB04a}. 
These \textit{synthetic antiferromagnets}
are of great interest in modern nanomagnetism,
in particular due to their application in 
spin electronics \cite{Kim01} and 
high-density recording technologies
\cite{FullertonIEEE03}.

In view of their magnetic states and 
field-induced reorientation transitions,
these antiferromagnetically coupled
superlattices can be separated 
into two groups:
%
(1) Systems with magnetization 
in the film plane 
and low (higher-order) anisotropies only, 
e.g., multilayers grown on (001) faces of cubic substrates
\cite{AFM001}  with a four-fold anisotropy 
owing to the magneto-crystalline anisotropies
of the materials (for further references 
and a survey of their magnetic properties, see \cite{PRB04a}).
In these high symmetry systems, 
magnetic states are mainly determined under competing
influence of bilinear and biquadratic exchange 
interactions and the intrinsic magnetic anisotropy.
For this type of multilayers
with a fully compensated 
antiferromagnetic collinear ground state,
the magnetization processes 
generally have a simple character, 
in particular in the low anisotropy limit 
no reorientation transition occurs 
with fields in direction of easy axes \cite{PRB04a}. 
(2) The other group of the synthetic antiferromagnets
own an often sizeable \textit{uniaxial} anisotropy. 
This group includes superlattices with 
intrinsic or \textit{induced} in-plane 
uniaxial anisotropy, e.g. multilayers on (110) and (211) 
faces of cubic substrates \cite{Wang94, Felcher02},
or nanostructures with perpendicular anisotropy
\cite{Hellwig03}.
Here, an  interplay between
the uniaxial anisotropy and 
the confining geometry of the multilayers 
determines their magnetic properties
and gives rise to effects such as \textit{surface spin-flops}
\cite{Wang94, Felcher02},
field induced cascades of magnetization jumps
\cite{Hellwig03, JMMM04} and
multidomain structures \cite{Nagy02, PRB04a}.

\begin{figure}[thb]
\includegraphics[width=6cm]{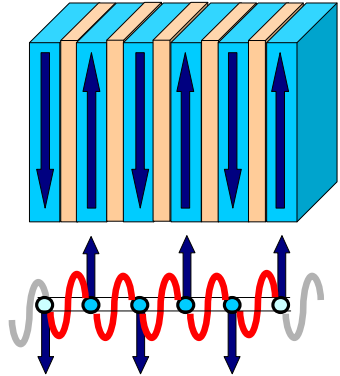}
\caption{%
\label{StackCartoon}
The magnetic states of an antiferromagnetic
superlattice stack can be represented 
by a linear chain of unity spins;
the ``exchange springs'' are cut at the ends 
of the finite chain.
}
\end{figure}

Theoretical investigations
of such systems have a long history \cite{Mills68}
and a large number of results on the magnetic
states (mostly within simplified models and 
by numerical methods) have 
been obtained \cite{Mills68,Trallori94,Mills99}. 
However, these findings were often restricted 
to specific values of magnetic parameters 
and led to conflicting conclusions about
the evolution of the magnetic states in these system.
In particular, the understanding of 
the phenomena called \textit{surface spin-flop}
remained controversial \cite{Trallori94}.
Recently, 
we have achieved a complete overview
on all one-dimensional solutions 
for the basic phenomenological model of 
antiferromagnetic superlattices with 
even number of layers $N$ \cite{PRB04},
and the magnetization processes 
in the limits of zero  \cite{PRB04a} 
and strong  \cite{JMMM04} uniaxial 
anisotropy have been investigated.
In this paper, we systematize and discuss
the results of \cite{PRB04,PRB04a, JMMM04, SFNew} 
giving a consistent description of magnetic 
states and their evolution in an applied magnetic field. 
Magnetization processes have been investigated 
in detail and corresponding phase diagrams are derived.
These reveal the qualitatively different behaviour
arising for low, high, and intermediate uniaxial anisotropy.
A ``dimerization'' transformation 
for the energy of 
an antiferromagnetic superlattice with $N$ layers 
into a chain of $N/2$ interacting ``antiferromagnets''
explains the physics ruling the magnetic phases 
in these nanostructures in the low anisotropy limit.
The approach elucidates 
the crucial role of the \textit{cut exchange}
for the formation of magnetic states 
in an antiferromagnetic layer \cite{PRB03}
and for the structure of the possible phase-diagrams
in finite antiferromagnetic multilayer stacks \cite{PRB04}.
At high anisotropies, the phase diagrams
display a fixed sequence of metamagnetic transitions
between collinear magnetic configurations.
This simplicity of the phase structure
of this micromagnetic model in the limiting cases 
allows to understand the general behaviour 
of its phase diagrams.
Our results resolve the puzzle of 
reorientation transitions in these systems, 
which have been discussed 
as ``surface spin-flop'' for many years.

\textbf{2. General model and simplifications. }

Following \cite{PRB04, PRB04a}, 
an antiferromagnetic superlattice is described
by a stack of $N$ ferromagnetic plates 
with magnetizations $\mathbf{m}_i$, 
antiferromagnetic couplings, and $N$ even.
To calculate the one-dimensional configurations,
one can replace this by 
a linear chain of coupled unity vector spins
${\mathbf s}_i={\mathbf m}_i/|{\mathbf m}_i|$.
For the system with uniaxial magnetic anisotropy
its phenomenological energy can be written as
\begin{eqnarray}
\Phi_N=\sum_{i=1}^{N-1} \left[ J_i\,\mathbf{s}_i \cdot
\mathbf{s}_{i+1}
+ \widetilde{J}_i\,\left(\mathbf{s}_i \cdot
\mathbf{s}_{i+1} \right)^2 \right]
-\mathbf{H}\cdot \sum_{i=1}^{N} \mathbf{s}_i
\nonumber\\
- \frac{1}{2}\sum_{i=1}^{N} K_i\left( \mathbf{s}_i\cdot 
\mathbf{n} \right)^2
 - \sum_{i=1}^{N-1} K'_i\left( \mathbf{s}_i\cdot 
\mathbf{n} \right)
\left( \mathbf{s}_{i+1}\cdot \mathbf{n} \right). 
\label{energy1}
\end{eqnarray}
Here, 
$J_i$ and $\tilde{J}_i$ are bilinear 
and biquadratic exchange constants, respectively.
The unity vector $\mathbf{n}$  points
along the common anisotropy axis;
$K_i$ and $K'_i$ are constants of in-plane 
and inter-planar anisotropy. 
%

%
\begin{figure}[t]
\includegraphics[width=6cm]{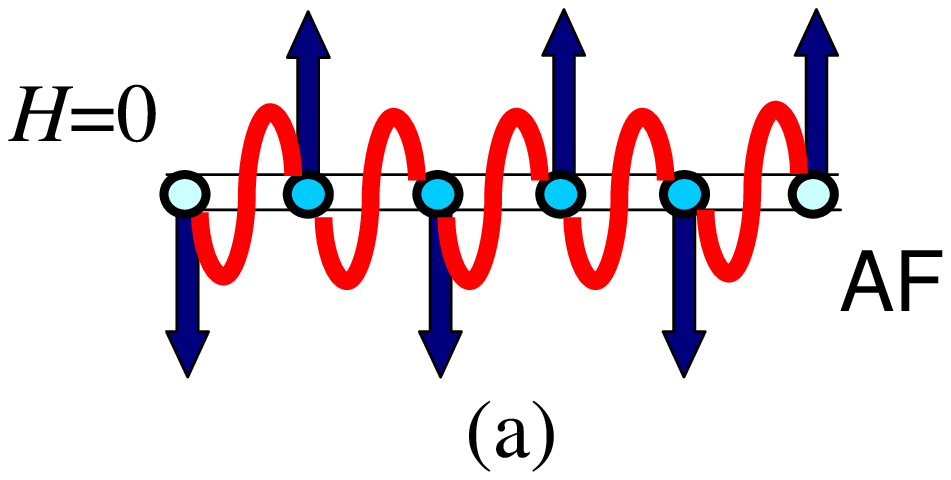}\hspace{1.85cm}
\includegraphics[width=6cm]{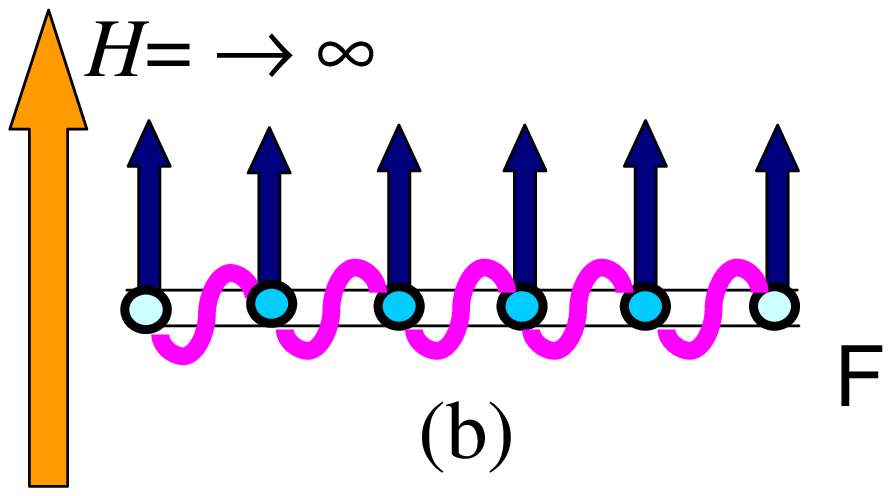}\hspace{1.75cm}
\includegraphics[width=6cm]{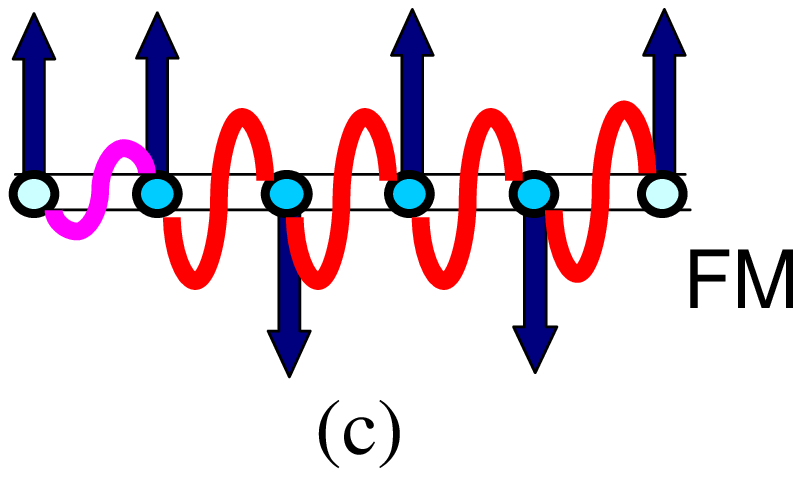}
\caption{%
\label{States}
Basic configurations: (a) antiferromagnetic ground state,
(b) saturated state in large fields,
and (c) ferrimagnetic states (FM) 
for large anisotropy at intermediate fields.
}
\end{figure}

For a detailed discussion of the model (\ref{energy1}),
its limits and relations to other theories,
see \cite{PRB04a, SFNew}.
The functional (\ref{energy1}) includes 
all main energy contributions which are
known to play a noticeable role in 
these nanostructured systems. 
It can be used for a detailed description 
of specific systems and for 
the analysis of experimental data.
Among the different magnetic interactions
in (\ref{energy1}), 
there are two factors which cause 
the striking behaviour of 
this class of magnetic nanostructures:
(i) \textit{uniaxial} anisotropy 
and 
(ii)  \textit{cut exchange bonds} \cite{PRB04}.
Anisotropy is mandatory for 
field-driven reorientation.
The second effect is illustrated 
in Fig.~\ref{StackCartoon}:
interior layers are coupled to two neighbours,
while the endmost layers interact only with one. 
The corresponding weakening of the exchange stiffness 
at the ends of the chain, 
the \textit{cut exchange bonds} at the boundary layers,
determines the reorientational effects.
In principle, the same effect of cut exchange bonds 
rules the magnetization processes of a geometrically 
confined antiferromagnet with non-compensated surfaces.
Similar effects cannot arise in \textit{ferromagnetically} 
coupled multilayers (or ferromagnetic films) because 
here reorientational processes do not depend on 
the relative strength of the exchange interaction 
between the layers at the surface or 
in the interior.

The effects of uniaxial anisotropy and 
the exchange cut can be described
by a simplified version of the model (\ref{energy1}) 
with $J_i=J$, $K_i=K$, and $\widetilde{J}_i = K'_i =0 $ 
for $i=1$ to $N$.
The vectors $\mathbf{s}_i$ 
can be confined to a plane, which includes 
the easy axis $\mathbf{n}$. 
For an applied field in direction of the easy axes,
$\mathbf{H} \parallel \mathbf{n}$, the energy reads
\begin{eqnarray}
\widetilde{\Phi}_N  = 
J\sum_{i=1}^{N-1} \cos(\theta_i - \theta_{i+1})
-H \sum_{i=1}^{N}\cos\theta_i
- \frac{K}{2}\sum_{i=1}^{N} \cos^2 \theta_i\,,
\label{energyMills}
\end{eqnarray}
where $\theta_i$ is the angle between $\mathbf{s}_i$ and
$\mathbf{n}$, and for $K > 0$ the axis $\mathbf{n}$
is the easy direction of the magnetization.

The model (\ref{energyMills}) for the magnetic 
energy was introduced by Mills 
for a semi-infinite chain \cite{Mills68},
and it has been studied in many works 
(see \cite{Trallori94} and  bibliography in \cite{SFNew}).
This model, called here \textit{Mills model},
is a basic model to discuss magnetic properties
of antiferromagnetic superlattices.
In the following sections we consider the
structure of solutions for (\ref{energyMills}).

\textbf{3. Special cases of Mills model. }

In Eq. (\ref{energyMills}) there are
three independent control parameters:
number of the layers  $N$, and the ratios $H/J$ and $K/J$.
We represent this phase space of the model 
by ($H/J$, $K/J$) diagrams for different values of $N$.
We start the analysis of Mills model from
the limits of the control parameters, 
where the system becomes simple.

\textit{Limiting values of the field $H$.}
In these phase diagrams the (zero-field)
ground state is always the \textit{antiferromagnetic}
phase (AF) with $\theta_{2i-1} = \pi$
and $\theta_{2i} = 0$ (Fig.~\ref{States}(a)).
At high field ($H \rightarrow \infty$) the minimum of
the energy (\ref{energyMills}) corresponds to
the \textit{ferromagnetic} state (F) with $\theta_i = 0$ 
(Fig.~\ref{States}(b)).
\begin{figure}[thb]
\includegraphics[width=10cm]{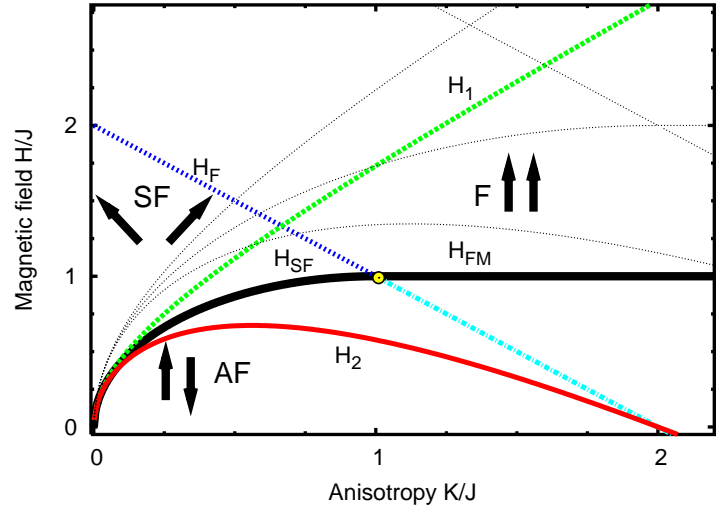}
\caption{%
\label{N2PD}
Phase diagram for two antiferromagnetically 
coupled uniaxial layers $N=2$, or 
equivalently a two-sublattice 
bulk antiferromagnet \cite{Neel36},
with field along easy axis.
Light grey lines give 
transition lines,
rescaled by $J\rightarrow 2J$,
as approximation for phase diagrams 
of multilayer systems $N\rightarrow\infty$.
}
\end{figure}

\textit{The two limits of $N$.}
For the simplest antiferromagnetic layer $N = 2$, 
the energy (\ref{energyMills})
reduces to that of a classical 
bulk two-sublattice antiferromagnet
\begin{eqnarray}
\widetilde{\Phi}_2  = 
J \cos(\theta_1 - \theta_2)
-H (\cos\theta_1 + \cos\theta_2)
- \frac{K}{2}(\cos^2 \theta_1 + \cos^2 \theta_2).
\label{energyMills2}
\end{eqnarray}

The corresponding phase diagram is plotted in Fig.~\ref{N2PD} .
For $K < J$ a first-order transition between
AF ($\theta_1 = 0$, $\theta_2 = \pi$) and
the \textit{spin-flop} (SF) ($\theta_1 = -\theta_2)$
phase occurs at $H_{\mathrm{SF}} = \sqrt{K(2J-K)}$.
This is the \textit{spin-flop} transition \cite{Neel36}.
In the SF phase $\cos \theta_1 = H/H_{\mathrm{F}}$,
where $H_{\mathrm{F}} = 2J-K$ is the field of 
the continuous (second-order) transition 
into the ferromagnetic phase.
The critical fields $H_1 = \sqrt{K(2J+K)}$,
$H_2 = H_{\mathrm{F}} \sqrt{K/(2J+K)}$ are stability limits
for the AF and SF phase, correspondingly.
For $K > J$ the first-order transition between
AF and ferromagnetic phases
(\textit{metamagnetic} phase transition) occurs
at the critical field $H_{\mathrm{FM}}= J$. 
For this transition
the continuation of the line $H_{\mathrm{F}}(K)$ 
gives the stability limit of the ferromagnetic phase.
In multilayers with $N > 2$ the exchange coupling
per any internal layer is $2J$. 
Thus, the phase diagram  for the systems 
with very large $N$ 
can be obtained from that for $N = 2$ 
by substitution $J \rightarrow 2J$. 
These critical lines are shown in Fig.~\ref{N2PD}.
This procedure does not yield 
a completely correct phase diagram 
because one neglects the cut of exchange bonds 
at the boundaries. 

\begin{figure}[thb]
\includegraphics[width=8cm]{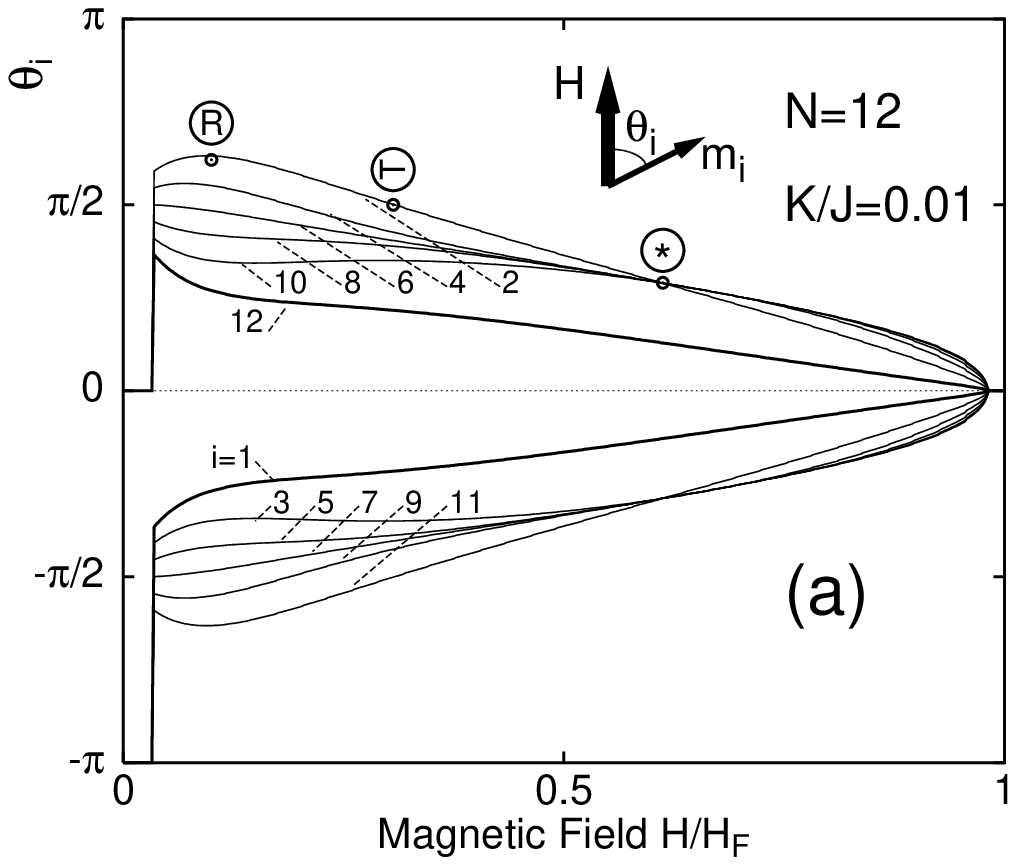}
\includegraphics[width=8cm]{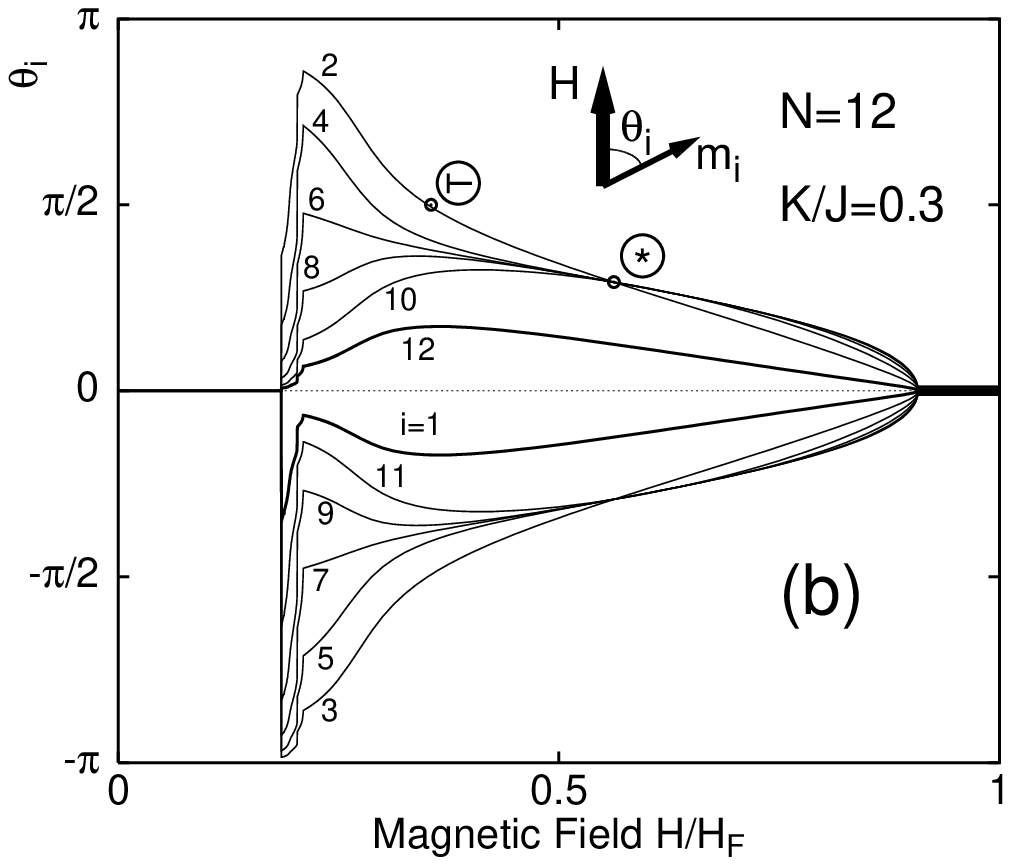}
\caption{%
\label{ThetaN12}
Evolution of configuration profiles
for Mills model $N=12$ with low anisotropy 
(a) and for intermediate anisotropies (b) 
with field along the easy axis.
In the SF-phase in (a) some moments rotate against 
the external field, e.g. moment~2, up to point (R);
at ($\vdash$) it is perpendicular 
to the field. 
At point ($\ast$), 
projections of all interior moments 
onto the field are equal. 
With stronger anisotropy, a series of 
canted and asymmetric configurations occurs
between AF and SF-state. 
}
\end{figure}

\textit{High and low anisotropy}.
Sufficiently large anisotropy suppresses
deviations of the magnetization from the easy directions.
In this limit, only phase transitions 
between the collinear
states are possible \cite{JMMM04}. 
In particular for Mills model there are two
phase transitions \cite{Trallori94, PRB03}:
at $H_{\mathrm{FM}} = J$ between AF and 
\textit{ferrimagnetic} (FM) phases 
with a flipped pair of moments (Fig.~\ref{States}(c)) 
and  at $H_{\mathrm{FM2}} = 2J$ 
between FM and the ferromagnetic state. 
The transition from the AF-phase 
into the collinear FM states
is again a consequence of the exchange cut.
The moment at the surface 
pointing against the field 
can be reversed more easily  
than a moment in the interior.
However, the realizations of 
the ferrimagnetic phases are degenerate
for Mills model.
They are built from one pair of
ferromagnetically aligned spins $\uparrow\uparrow$
between two collinear antiferromagnetic domains,
so that the two endmost moments 
point in direction of the field:
$(\uparrow\downarrow\dots\uparrow\downarrow)$
$\mathbf{\uparrow\uparrow}$ $(\downarrow\uparrow\dots\downarrow\uparrow)$.
All configurations with different location 
of the ferromagnetic pair, equivalently with different 
lengths of the two adjoining antiferromagnetic domains, 
have the same energy for Mills model Eq.~(\ref{energyMills}).
Therefore, they have 
the same transition fields $H_{\mathrm{FM}}$
and $H_{\mathrm{FM2}}$ 
for the first-order transitions 
into the antiferromagnetic 
ground state  and
into the ferromagnetic phase, respectively.
However, their stability regions are 
different and depend on $N$.
This remarkable degeneracy is due to
the highly symmetric choice for 
the materials constants of individual layers 
in energy (\ref{energyMills}).
It does not hold for general
cases described by Eq.~(\ref{energy1}) \cite{JMMM04}.

Next, we consider the opposite limit of
low anisotropy. 
Due to the \textit{cutting of exchange bonds} 
the SF phase in the systems with $N > 2$ 
has an inhomogeneous structure
across the stack and it undergoes a 
complex evolution 
in an applied magnetic field \cite{PRB04, PRB04a}.
An example is shown in Fig.~\ref{ThetaN12}(a).
For the highly symmetric Mills model 
this spin-flop state preserves mirror symmetry 
about the center of the layer, $\theta_i=-\theta_{N+1-i}$.
For stronger anisotropies, such an inhomogeneous 
spin-flop state is reached only at higher fields, 
see Fig.~\ref{ThetaN12}(b).

\textbf{4. $H$--$K$ phase diagram for Mills model.}

\begin{figure}[bth]
\includegraphics[width=16cm]{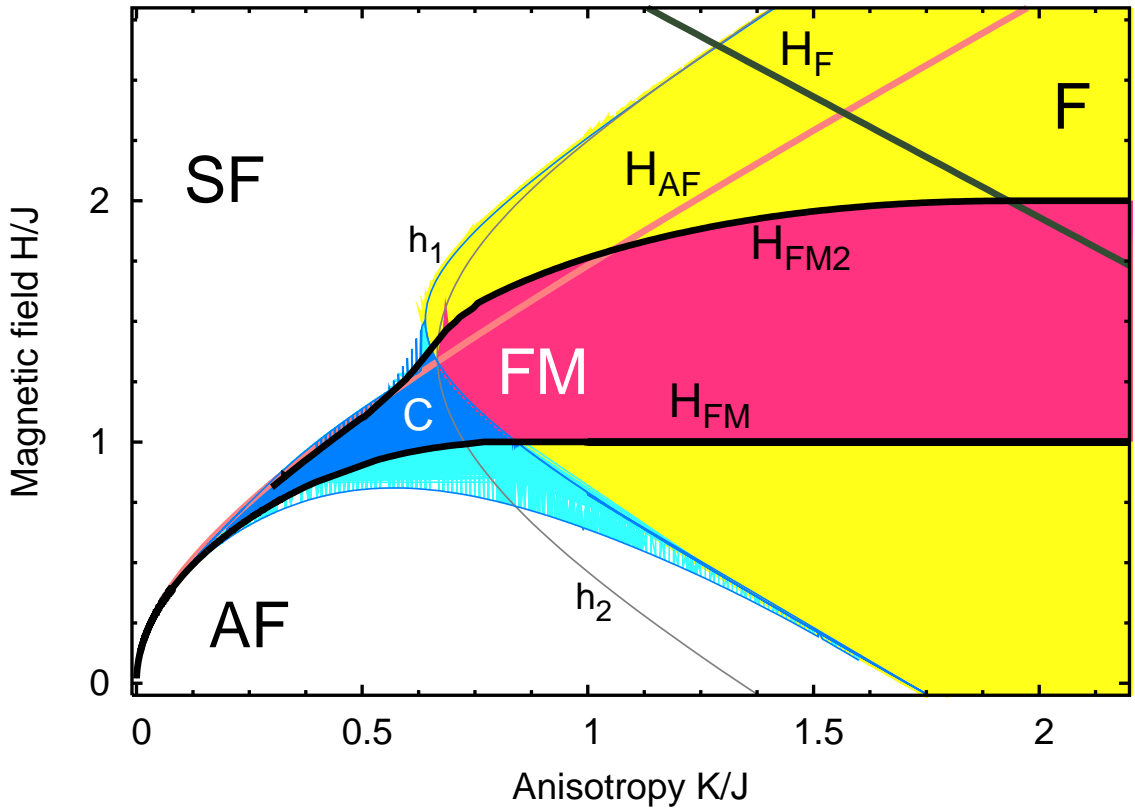}
\caption{%
\label{PD12}
Phase diagram 
Mills model Eq.~(\ref{energyMills})
with $N=12$ layers 
representative for 
phase structures of uniaxial 
antiferromagnetic superlattices
with field along easy axis.
The collinear ferrimagnetic states 
are stable in the shaded region FM. 
First-order transitions 
occur at lines $H_{\mathrm{FM}}$ and $H_{\mathrm{FM2}}$.
Lines $h_1$ and $h_2$ are stability 
limits of different FM-phases.
Canted asymmetric 
phases exist in region C,
which is grossly simplified here.
Lighter shaded areas mark 
metastability limits of FM- and C-phases. 
Line $H_{\mathrm{F}}$: continuous transition 
between spin-flop and saturated phase,
and stability limit of F below 
line $H_{\mathrm{FM2}}$.
}
\end{figure}

\begin{figure}[thb]
\includegraphics[width=16cm]{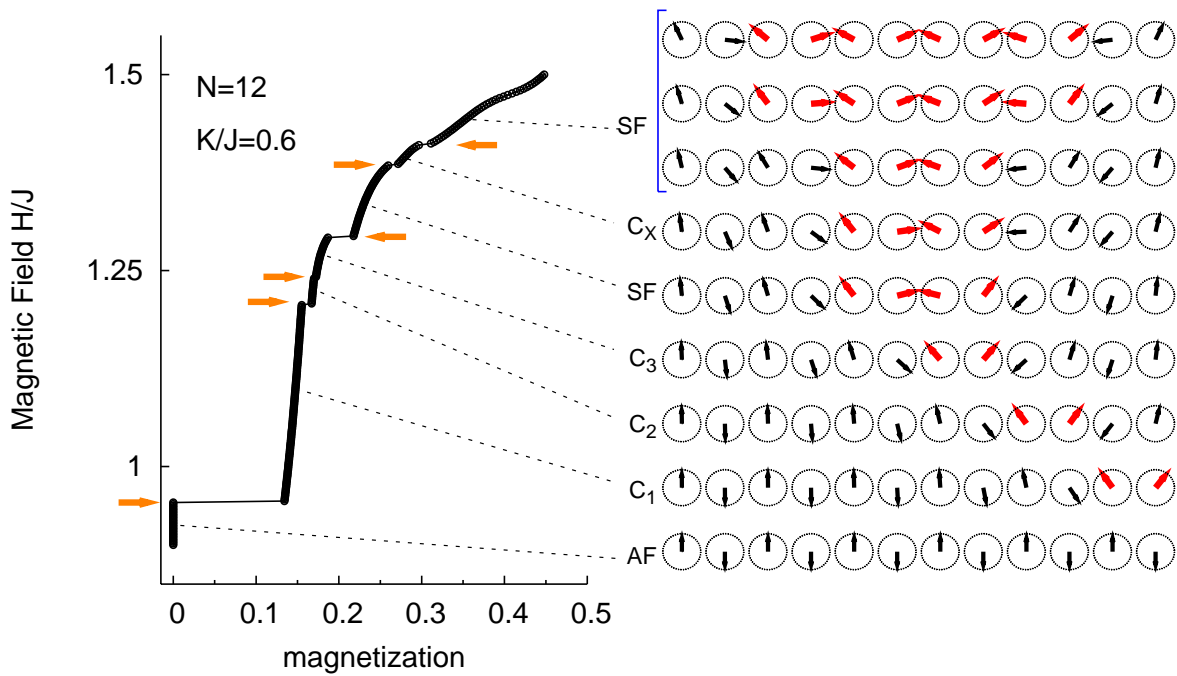}
\caption{%
\label{EvolN12}
Evolution of equilibrium magnetic states 
for Mills model with $N=12$ layers 
in the region of canted phases.
From bottom to top a series of 
transition leads from antiferromagnetic 
to spin-flop phase with increasing field.
Left panel: magnetization 
(first-order transitions 
are marked by arrows).
Right panel: corresponding configurations.
}
\end{figure}

\begin{figure}[thb]
\includegraphics[width=10cm]{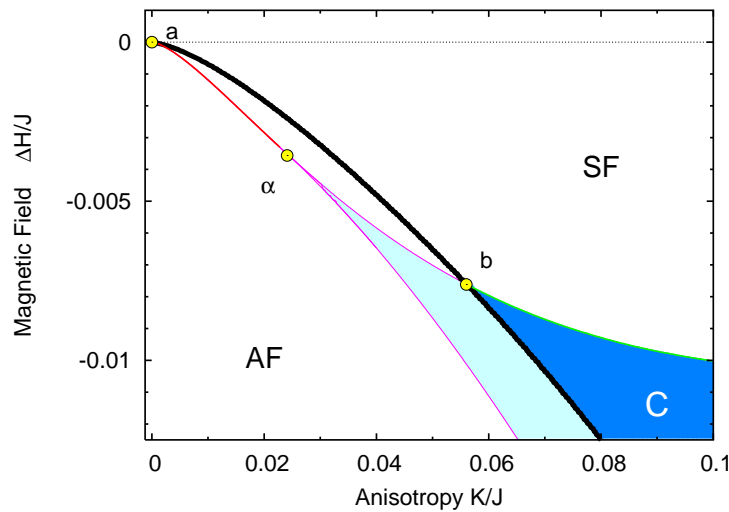}
\caption{%
\label{PD12lowK}
Low-anisotropy and low-field part of 
the $N=12$ phase diagram in Fig.~\ref{PD12}.
Magnetic field is given 
relative to the metastability field of 
AF-phase $\Delta H=H_{\mathrm{AF}}-H$.
First-order transitions take place 
at the thick black line: 
from AF to SF-phase between points a and b;
for larger anisotropy
beyond point b, 
from AF to C$_1$-phase (see Fig.~\ref{EvolN12}).
In the dark area the canted phase C$_1$ is stable,
the lighter shaded area is its metastability range
beyond the first-order transition line.
Point $\alpha$ is the low anisotropy limit,
where the C$_1$-phase is metastable.
}
\end{figure}

A phase diagram for an antiferromagnetic superlattice
is shown in Fig.~\ref{PD12} as representative for
all cases.
At large anisotropies, 
one has the simple sequence of collinear 
phases AF $\rightarrow$ FM $\rightarrow$ F
separated by first-order (metamagnetic) transitions.
The special line $H_{\mathrm{AF}}$ 
is the stability limit of the AF-phase.
For Mills model, one has $H_{\mathrm{AF}}=\sqrt{2JK+K^2}$
for all $N \ge 4$ and even \cite{Trallori94}.
In Fig.~\ref{PD12}
stability limits for two realizations of the FM-phases are shown: 
$h_1$ for phase FM$_1$ $=(\uparrow\downarrow\dots)\uparrow\uparrow$
with the ferromagnetic pair at the surface, 
$h_2$ for FM$_2$ 
$=(\uparrow\downarrow\dots)\uparrow\uparrow(\downarrow\uparrow)$.
Thus, at intermediate anisotropies the collinear FM-phases 
can be distorted elastically. 
These are canting instabilities. 
The transitions between the collinear phase FM$_i$ 
and a corresponding canted phase C$_i$ 
usually are continuous. 
These instabilities lead to the appearance
of series of different asymmetric phases
in the region marked ``C'' in Figs.~\ref{PD12} 
and \ref{PD12lowK}.
A typical evolution through this region 
is demonstrated in Fig.~\ref{EvolN12},
there are series of first-order transitions 
between various canted phases.
The low anisotropy and low-field part of the
phase diagram in Fig.~\ref{PD12lowK} 
is shown with respect to 
the stability limit $H_{\mathrm{AF}}$. 
For small anisotropies below point b,
there is only 
the spin-flop transition AF $\rightarrow$ SF.
Finally, the SF-phase reaches the 
saturated F-state continuously at 
the straight line $H_{\mathrm{F}}=H_e^{(N)}-K$
with an exchange field $H_e$ depending on $N$
(Fig.~\ref{PD12}).

\textbf{5. Does a surface spin-flop occur in antiferromagnetically
coupled multilayers?}

a. \textit{Recent picture of a \emph{surface spin-flop.}}
A common scenario 
for reorientational transitions in 
antiferromagnetic multilayers 
(see \cite{Mills99,Felcher02})
was built on the basis of analytical 
investigations for low anisotropy systems \cite{Mills68}
and numerical simulations for systems with
higher anisotropy \cite{Wang94,Trallori94}.
It is described as a local instability 
driven by the field at the surface of the
uniaxial antiferromagnet with a moment 
pointing against the direction of the field. 
This should create a flopped
configuration at this surface,
hence, an asymmetric state.
This picture seems to be born out by the evolution 
through sequences of canted phases 
as shown in Fig.~\ref{EvolN12} for systems 
with sizeable anisotropy. 
But, it is not completely correct.
The transition field for the surface spin-flop was 
identified with the instability field of 
the antiferromagnetic phase 
$H_{\mathrm{AF}}\simeq\sqrt{2JK}$ \cite{Wang94,Mills68}.
With increasing field, 
this flopped configuration ``moves'' 
into the center of the stack
and finally the flopped configurations 
should expand by another transition 
at the field for 
the bulk spin-flop $H_{\mathrm{B}}\simeq\sqrt{4JK}$.  
Therefore, the ratio $\sqrt{2}$ 
between the field for this ``surface spin-flop'' and 
the bulk spin-flop should hold.
The first-order transition AF $\rightarrow$ C$_1$ 
clearly resembles an isolated flop at the surface.
However, it has to take place considerably 
below $H_{\mathrm{AF}}$ (see Fig.~\ref{EvolN12}).
The anomalies identified 
with the ``bulk spin-flop'' are related 
to complicated canting instabilities
at fields below $H_{\mathrm{B}}$ 
(transitions between SF and C$_x$ in Fig.~\ref{EvolN12}).

b. \textit{The spin-flop in the low anisotropy limit.}
The picture of the ``surface spin-flop'' 
also is incomplete. 
There are no asymmetric phases 
or isolated flopped configurations 
at the surface of a multilayer with low anisotropy 
(Fig.~\ref{PD12lowK}).

For low anisotropy systems ($K << J$) and weak magnetic
fields ($H << J$) the energy (\ref{energyMills}) can be
reduced to a model which allows to extract the physical
mechanism behind the complex and unusual reorientational 
processes described in the previous sections.
Following a standard procedure we introduce
for a two-layer system  with (\ref{energyMills2})
the vector of the net magnetization
$\mathbf{M}= (\mathbf{s}_1 + \mathbf{s}_2)/2$
and staggered vectors
$\mathbf{L}= (\mathbf{s}_1 - \mathbf{s}_2)$.
for each antiferromagnetic pair.
From $\mathbf{s}_{i}$ = 1 follows
that $M^2 + L^2 = 1$
and $\mathbf{M} \cdot \mathbf{L} $ =0.
Using these equations the energy (\ref{energyMills2})
can be written as a function of $|\mathbf{M}|$
and the angle $\phi$ between $\mathbf{L}$
and the easy axis. 
After minimization with respect
to $|\mathbf{M}|$ the energy can be reduced to
the following form
\begin{eqnarray}
\widetilde{\Phi}_2  = 
 - \frac{H^2 - 2JK}{4J} \cos 2 \phi\,.
\label{energyMills2a}
\end{eqnarray}
Compared to the energy ({\ref{energyMills2})
the Eq.~(\ref{energyMills2a}) includes
only leading contributions with respect 
to the small parameters $K/J$, $H/J$. 
In this limit
the spin-flop field and 
lability fields of AF and SF phases become
equal ($H_{\mathrm{SF}}=H_1 = H_2= \sqrt{2JK}$)
This equation clearly demonstrates 
the competing character of the magnetic states in a bulk
two-sublattice easy-axis antiferromagnet
and the physical essence of the spin-flop transition.
The uniaxial anisotropy stabilizes the AF phase
with the staggered vector along the easy-axis 
while the applied field favours the flopped states
with $\mathbf{L} \bot \mathbf{n}$.
At the threshold \textit{spin-flop} field
$H_{\mathrm{SF}} = \sqrt{2JK}$ the system switches from 
one mode to another.

Now we simplify the energy of the superlattice
with $N$ layers considering it as a system
of interacting \textit{dimers}.
We sort $N$ magnetic moments in the chain into $N/2$ pairs:
($\mathbf{s}_1, \mathbf{s}_2$),
($\mathbf{s}_3, \mathbf{s}_4$),...,
($\mathbf{s}_{2j-1}, \mathbf{s}_{2j}$),...,
($\mathbf{s}_{N-1}, \mathbf{s}_{N}$) with $j = 1,..., N/2$.
The chain of spins for the magnetic moments
then appears as a chain of \textit{dimers},
and each of these can be considered
as a two-sublattice antiferromagnet.
Net magnetization
$\mathbf{M}_j= (\mathbf{s}_{2j-1} + \mathbf{s}_{2j})/2$
and staggered vectors
$\mathbf{L}_j= (\mathbf{s}_{2j-1} - \mathbf{s}_{2j})/2$
are introduced for each antiferromagnetic pair;
from $\mathbf{s}_{i}$ = 1, one now has 
$M_j^2 + L_j^2 = 1$ and $\mathbf{M}_j \cdot \mathbf{L}_j $ =0.
For $H,K << J$  the net magnetization
is always small, $\mathbf{M}_j << 1$.
Thus, after expanding (\ref{energyMills}),
one can use an independent minimization
with respect to the variables $\mathbf{M}_j$.
\begin{eqnarray}
\widetilde{\Phi}_N  = 
 \frac{J}{2} \sum_{j=1}^{N/2-1} (\phi_{j+1} - \phi_{j})^2
 - \sum_{j=1}^{N/2}\frac{H^2 - \Lambda_j K}{2\Lambda_j} \cos 2 \phi_j\
 + \Xi( \phi_j)\,.
\label{energyMills3}
\end{eqnarray}
From the minimization 
the net magnetizations are fixed 
as dependent parameters 
by $M_j = H \sin \phi_j/\Lambda_j$.
The constants are $\Lambda_1$ = $\Lambda_{N/2}$ = $3J$ 
for the pairs at the surfaces, and 
$\Lambda_j = 4J$  
for the interior of the stack, $j= 2,..., N/2-1$.
The last contribution in (\ref{energyMills3}) 
is a ``surface''-term given by
\begin{eqnarray}
\Xi( \phi_j) =
-J\left( \Lambda_2^{-1} + \Lambda_1^{-1}\right)
\left(\mathbf{H} \cdot \mathbf{l}_1
- \mathbf{H} \cdot \mathbf{l}_{N/2} \right)
-J\left( \Lambda_2^{-1} - \Lambda_1^{-1}\right)
\left(\mathbf{H} \cdot \mathbf{l}_2
- \mathbf{H} \cdot \mathbf{l}_{N/2-1} \right)\,,
\label{energyXi}
\end{eqnarray}
where $\mathbf{l}_j = \mathbf{L}_j/|\mathbf{L}_j|$
are unity vectors along the staggered magnetization.
The energy contribution (\ref{energyXi}) is 
explicitly given by 
$\Xi( \phi_j)=-7H(\cos \phi_1 - \cos \phi_{N/2})/12
+H(\cos \phi_2 - \cos \phi_{N/2-1})/12$.
Comparison between the energies
$\widetilde{\Phi}_2$ (\ref{energyMills2a}) and
$\widetilde{\Phi}_N$  (\ref{energyMills3})
helps to understand 
the different character of reorientational transitions 
in the antiferromagnetic chain.

Eq.~(\ref{energyMills3})  can be considered
as the energy of the interacting \textit{dimers}
with "self-energy" (\ref{energyMills2a}).
The first term in (\ref{energyMills3}) has
the form of an elastic energy arising from
the exchange coupling between \textit{dimers}.
The second is a potential energy, which
changes its wells at critical values
of the field $H=\sqrt{\Lambda_j K}$.
There are two different critical fields: 
$H_{\mathrm{S}}=\sqrt{3 JK}$
for the surfaces, while in the interior 
one has $H_{\mathrm{B}}=\sqrt{4 JK}$.
The two fields $H_{\mathrm{S}}$
and  $H_{\mathrm{B}}$ would correspond 
to independent spin-flops
of the antiferromagnetic pairs
either at the surface or in the bulk
in the absence of couplings along the chain.

Clearly, $H_{\mathrm{S}}$ is an upper limit
for an instability of the AF-state in a finite superlattice.
It would correspond to a ``true surface spin-flop''.
But the transition from AF to SF
is driven by the energy contribution 
from the antisymmetric ``surface''-terms
$\Xi$ Eq.~(\ref{energyXi}), 
which describe the effect of the cut exchange.
This contribution becomes negative 
when the staggered vectors
at both ends of the chain, $\mathbf{l}_1$ 
and  $\mathbf{l}_{N/2}$, are antiparallel.
Thus, before reaching 
the field $H_{\mathrm{S}}$, 
a configuration resembling 
a 180-degree antiferromagnetic domain wall 
will be created in the chain.
The symmetry of the energy terms 
in (\ref{energyMills3}) only allows
(mirror) symmetric configurations. 
These are the 
symmetric SF-state (Fig~\ref{ThetaN12}(a)).
In multilayers with large $N$ 
the SF-configuration 
may appear as a rather localized wall-like structure
in the center of the system between antiferromagnetic
domains with antiparallel staggered vectors.
In this state reaching the field $H_{\mathrm{B}}$, 
the antiferromagnetic configuration of 
the interior antiferromagnetic pairs must change.
Asymptotically with $N\rightarrow\infty$,
this is the approach to the classical spin-flop 
for the corresponding infinite 
bulk antiferromagnet (Fig.~\ref{N2PD}).

Thus, we have the following 
important conclusion.
The reorientational transitions 
in these confined antiferromagnets
known as \textit{surface spin-flop}
are not related to \textit{surface} states,
and they are no \textit{spin-flops}.
It is a transition between 
the AF and the inhomogeneous
spin-flop phase induced by
the instability of the
antiparallel magnetization in the
endmost layers owing to the exchange cut.
It is not related to the competition 
between the applied 
magnetic field and the uniaxial anisotropy
as in bulk antiferromagnets.
Owing to the elastic couplings of the system,
there is no magnetic transition
related to the ``bulk'' threshold field $H_{\mathrm{B}}$ 
for finite superlattices.
However a strong magnetic anomaly may 
be observable near $H_{\mathrm{B}}$. 

\textbf{6. Conclusions.}

We have demonstrated a fundamental 
difference of the spin-flop behaviour in multilayers 
with low anisotropy and with higher anisotropy.
At low anisotropy, 
there is only an inhomogeneous symmetric
spin-flop phase with a wall-like configuration 
in the center or somewhere below the surfaces.
This should also hold for films 
made of usual antiferromagnetic materials where $K/J<<1$. 
This result explains the failure
to observe a surface spin-flop 
as a state nucleated at surfaces of 
antiferromagnetic layers.
Canted asymmetric states, which show 
a flopped configuration at the surface, 
are observed in multilayers with larger 
anisotropies \cite{Wang94,Felcher02}.
Metamagnetic transitions 
are found in antiferromagnetic multilayers with strong 
perpendicular anisotropy\cite{Hellwig03}.
By an extension of our results 
for the generic Mills model, 
an analysis of the general models (\ref{energy1})
is feasible.
Their phase diagrams can be similarly partitioned 
into the three characteristic regions: 
at low anisotropies there is a sequence  
AF $\rightarrow$ SF $\rightarrow$ F-phase 
with an inhomogeneously distorted SF-phase;
at high anisotropies cascades of 
metamagnetic transitions between 
collinear phases take place \cite{JMMM04}. 
Number and structure of these
collinear phases determine 
the intermediate region of 
the phase space, where canting instabilities occur
and asymmetric transitional phases appear. 
In this region, the phase diagrams 
may still reflect some features 
described as ``surface spin-flop''\cite{Wang94}; 
however, new types of magnetic states 
may be stabilized and 
complicated magnetization processes will arise
when the particular degeneracy of Mills model is lifted.

\textbf{Acknowledgments}
A.\ N.\ B.\ thanks H.\ Eschrig for support and
hospitality at the IFW Dresden.

\end{document}